\documentclass[3p, preprint, times, 10pt]{elsarticle}

\usepackage{lineno}
\usepackage{chngpage}
\usepackage{float}
\usepackage{enumitem}
\usepackage{graphicx}
\usepackage{xcolor}
\usepackage[caption=false]{subfig} 

\usepackage{algorithm}
\usepackage[]{algpseudocode}
\usepackage{mathtools}
\usepackage{amssymb}

\usepackage[colorlinks=true, citecolor=blue, urlcolor=black, linkcolor=magenta]{hyperref}
\usepackage{seqsplit}
\usepackage{xurl}

\usepackage{comment} 
\usepackage{soul}

\usepackage{longtable} 
\usepackage{pdflscape,booktabs,tabularx,ragged2e}
\usepackage{multirow}
\newcolumntype{C}{>{\centering\arraybackslash}X}
\newcolumntype{L}{>{\raggedright\arraybackslash}X}
\newcolumntype{R}{>{\raggedleft\arraybackslash}X}

\usepackage{microtype} 
\usepackage{flushend}

\begin{document}
\microtypesetup{activate=true}

\begin{frontmatter}
\title{An Edge-Computing Based Industrial Gateway For Industry 4.0 Using ARM TrustZone Technology}
\author[label1]{Sandeep Gupta}\ead{sandeep.gupta@unitn.it}
\address[label1]{University of Trento, Italy}

\begin{abstract}
Secure and efficient communication to establish a seamless nexus between the five levels of a typical automation pyramid is paramount to Industry 4.0. Specifically, vertical and horizontal integration of these levels is an overarching requirement to accelerate productivity and improve operational activities. Vertical integration can improve visibility, flexibility, and productivity by connecting systems and applications. Horizontal integration can provide better collaboration and adaptability by connecting internal production facilities, multi-site operations, and third-party partners in a supply chain.
{\color{black}
In this paper, we propose an Edge-computing based Industrial Gateway for interfacing information technology and operational technology that can enable Industry 4.0 vertical and horizontal integration.
} 
Subsequently, we design and develop a working prototype to demonstrate a remote production-line maintenance use case with a strong focus on security aspects and the edge paradigm to bring computational resources and data storage closer to data sources.
\end{abstract}

\begin{keyword}
Industry 4.0 \sep Industrial Gateway \sep Edge Computing \sep TrustZone Technology \sep Security \sep Remote Maintenance
\end{keyword}

\end{frontmatter}

\section{Introduction}
Industry 4.0 (I4.0) is the fourth industrial revolution that aims at manufacturing/production processes automation, end-to-end digital value chain creation, and real-time production-line maintenance with the help of vertical and horizontal integration of the five levels of a traditional industrial pyramid shown in Figure~\ref{fig:AutomationPyramid}~\cite{civerchia2017industrial}. {\color{black}\emph{Level 0} to \emph{Level 4}, i.e., field-level, control-level, supervision-level, planning-level, and management-level in I4.0 is comparable to the production system, programmable logic controllers (PLC), supervisory control and data acquisition (SCADA), manufacturing execution system (MES), and enterprise resource planning (ERP) in industrial pyramid~\cite{cortes2020digital}.} Vertical integration requires coordination between systems and applications between levels for improving visibility, flexibility, and productivity. Whereas, horizontal integration requires coordination between internal production facilities, multi-site operations, and third-party partners in the supply chain for better collaboration and adaptability~\cite{vieira2020supply}.

Consequently, a system is needed that can acquire and process data from \emph{Level 1} and can provide bidirectional communication channels for interacting with higher \emph{Levels 2, 3, 4} and third-party partners to accomplish the vertical and horizontal integration requirements. It is requisite for the system to provide computing resources that can process the collected data locally and can support features, such as \textit{data filtration} (e.g., segregating the confidential and non-confidential production-line data), \textit{secure communication channels} (e.g., guarantee both the confidentiality and integrity of the information being requested and provided), and \textit{threat profiling capabilities} (e.g., concrete information for intrusion detection systems, attack tree generation, threat forensic, threat modeling).

Therefore, the use of the Edge-computing paradigm can be a viable technology to design Industrial Gateways (IGs) by bringing computational resources and data storage closer to data sources~\cite{gupta2022non}. With the Edge-computing capabilities, the IG can provide a scalable solution for implementing the aforementioned functionalities, i.e., data collection, data filtration, networking support, audit logs generation, and smart interfaces. That can be further exploited by higher levels or third-party partners for achieving better operational intelligence, real-time asset usage optimization, and predictive maintenance. 

\begin{figure}[!ht]
    \centering
    \includegraphics[width=.5\linewidth, keepaspectratio]{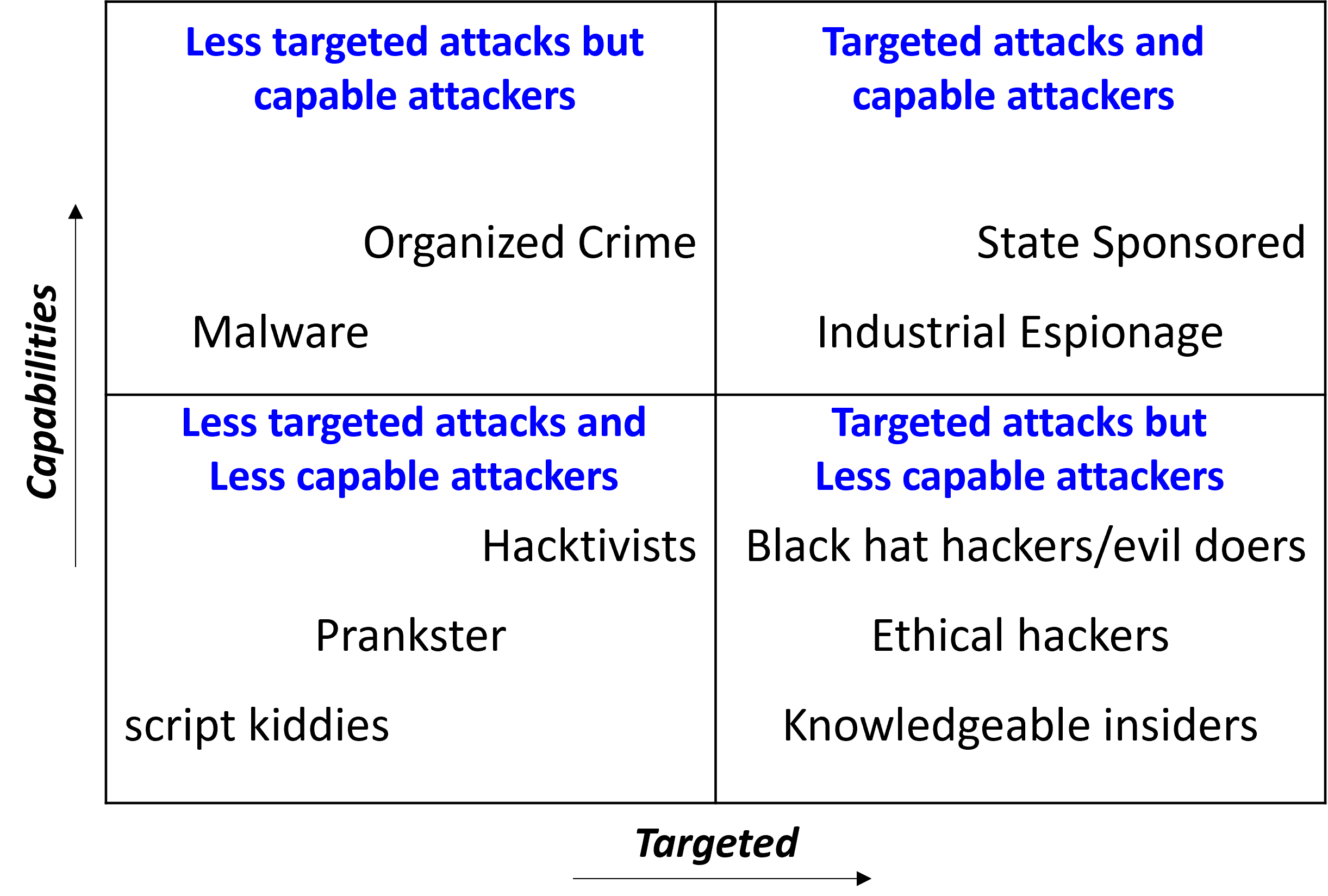}
    \caption{Adversaries motivation and capabilities}\label{fig:adversaries}
\end{figure}

However, interfacing Information Technology (i.e., information or data management) and Operation Technology (i.e., physical devices and processes) domains using internet technologies for horizontal and vertical integration can expose new cyber-attack surfaces. Figure~\ref{fig:adversaries} illustrates adversaries' intent to select a target, and their capabilities like skills and financial resources for cyber attacking an Industrial Internet of Things (IIoT) ecosystem~\cite{liebl2020threat}. Stuxnet, Saudi Aramco attack, cryptocurrency malware, Tridium Niagara framework attack, and Ukrainian Power Grid attack are some of the successful attacks in recent years that devastated critical operations~\cite{kayan2021cybersecurity}. Dhirani et al.~\cite{dhirani2021industrial} reported that 48\% of manufacturing industries have already suffered cyber attacks, however, only 11\% of manufacturers are willing to adopt adequate security measures and controls. Thus, security becomes the most critical non-functional requirement for preventing IGs from cyber-attacks including industrial espionage, organized crimes, denial of services, and data manipulation or disruption.   

In this paper, we design Edge-computing based Industrial Gateway (EC-IG) to bring computational resources and data storage closer to data sources for a seamless vertical and horizontal integration in I4.0 with a strong focus on security aspects. We investigate the design principles that can secure critical elements and minimize the possible attack surfaces. We exploit TrustZone technology for implementing a secure execution environment that can provide strong hardware-enforced separation to secure applications and data that is essential for critical sectors such as aerospace, automotive, and I4.0~\cite{pinto2019demystifying}. We discuss access control mechanisms for establishing the legitimacy of clients and requested services and boundary protection controls for inter-domain communications. We also introduce an intelligent threat profiling mechanism for audit trails and reviews.

The main contributions of this paper are listed below.
\begin{itemize}[leftmargin=*]
    \item Pilot study to perform remote production-line maintenance as a proof-of-concept.
    \item EC-IG architecture interfacing Level 0 and 1 with Level 2 to 4 for an I4.0.
    \item EC-IG prototype's high-level design and implementation on OP-TEE enabled QEMU targeting Armv8-A.
\end{itemize}

\textit{The rest of the paper is organized as}: Section~\ref{sec:Background} presents an overview of the industrial automation pyramid and ARM TrustZone technology. Section~\ref{sec:Relatedwork} covers the related work. Section~\ref{sec:PilotStudy} presents a pilot study and the threat model for a remote production-line maintenance use case. Section~\ref{sec:Architecture} describes the design principles and architecture of the proposed EC-IG prototype. Section~\ref{sec:Prototype} highlights the implementation details of the EC-IG prototype and the high-level design of the main features of the remote production-line maintenance. Section~\ref{sec:Discussion} presents a discussion on Trustzone assisted systems, access control mechanisms, boundary protection mechanisms, and 4) threat profiling. Finally, Section~\ref{sec:Conclusions} concludes the paper. 

Table~\ref{tab:Acronyms} presents the acronyms and their description used in this article.
\begin{table}[!ht]
    \centering
    \footnotesize
    \caption{Acronyms}\label{tab:Acronyms}
    \begin{tabularx}{.8\linewidth}{|l|L|}\hline
    \textbf{Acronyms} & \textbf{Description}\\\hline
    I4.0 & Industry 4.0 \\\hline
    I5.0 & Industry 5.0 \\\hline
    IIoT & Industrial Internet of Things \\\hline
    IG & Industrial Gateway\\\hline
    EC-IG & Edge-computing based Industrial Gateway\\\hline
    IT & Information Technology\\\hline
    OT & Operation Technology \\\hline
    PLC & Programmable logic controllers \\\hline
    SCADA & Supervisory control and data acquisition\\\hline
    MES & Manufacturing execution system\\\hline
    ERP & Enterprise resource planning\\\hline
    OPC & open communication standard\\\hline
    MQTT & Message Queuing Telemetry Transport\\\hline
    REST & Representational State Transfer\\\hline
    SNMP & Simple Network Management Protocol\\\hline
    SoC & System on Chip\\\hline
    CIA & Confidentiality, integrity, and availability\\\hline
    TEE & Trusted Execution Environments\\\hline
    REE & Rich Execution Environments\\\hline
    NW & Normal World\\\hline
    SW & Secure World\\\hline
    TA & Trusted application \\\hline
    SMC & Secure monitor call\\\hline
    IRQ & interrupts\\\hline
    FIQ & Fast interrupts\\\hline
    AXI & Advanced eXtensible Interface \\\hline
    STRIDE &  Spoofing, Tampering, Repudiation, Information disclosure, Denial of Service, and Elevation of privilege\\\hline
    \end{tabularx}
\end{table}

\section{Background}\label{sec:Background}
This section gives an overview of the industrial automation pyramid and Industrial Gateway generic requirements. Next, it discusses ARM TrustZone technology covering hardware architecture, hardware library, and software architecture.

\subsection{Industrial Automation Pyramid}
\begin{figure}[!ht]
    \centering
    \includegraphics[width=.6\linewidth, keepaspectratio]{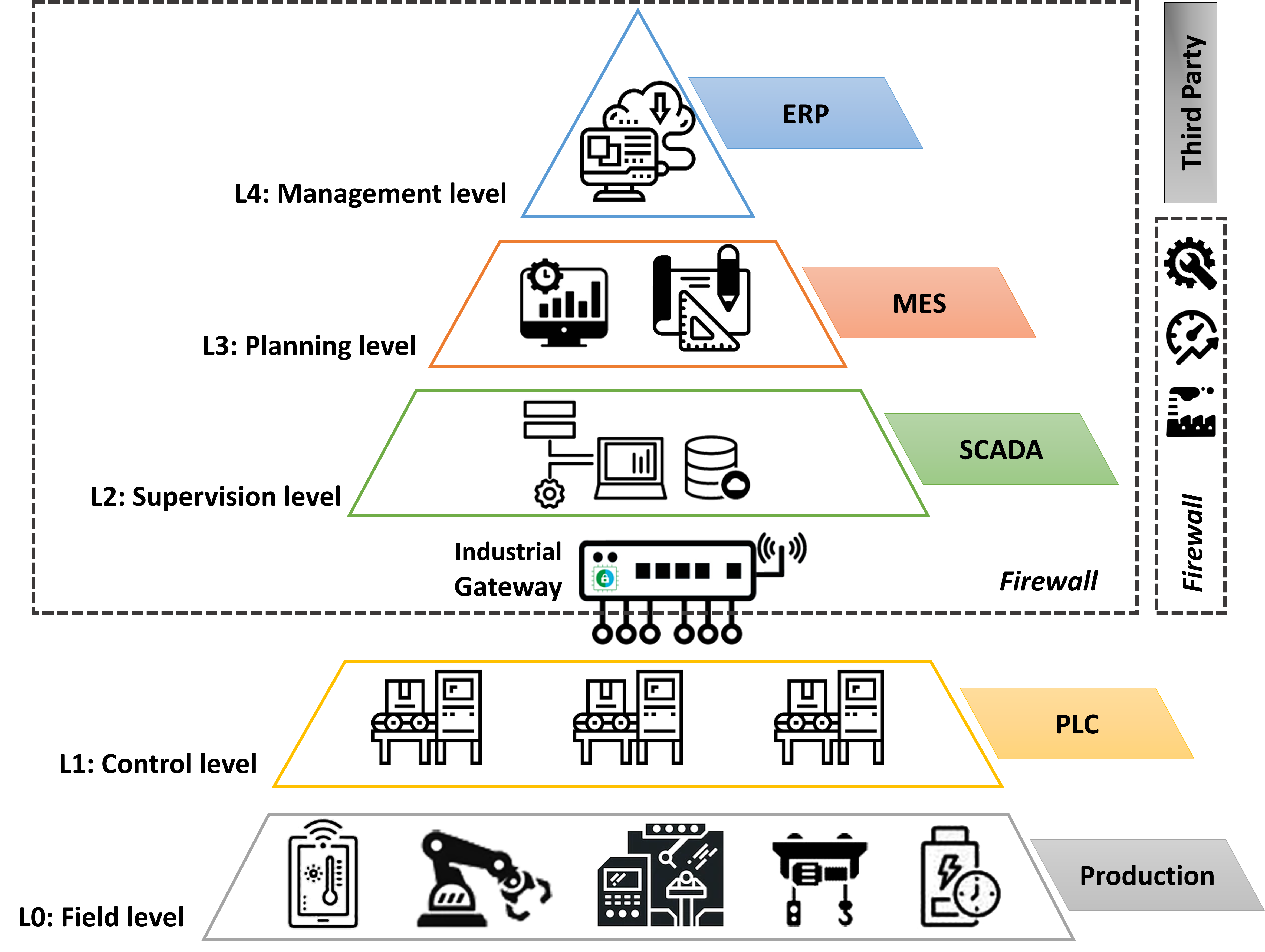}
    \caption{Industry 4.0 automation pyramid}\label{fig:AutomationPyramid}
\end{figure}

Figure~\ref{fig:AutomationPyramid} illustrates the industrial automation pyramid that consists of five levels~\cite{cortes2020digital}. IGs can be described as one of the pivotal systems for interfacing Level 1 with the higher levels and third parties to achieve industrial automation. 
Some of the technical requirements for IGs can be gathering data, filtering data, and supporting the standard industrial protocols (e.g., Modbus, ProfiNet, open communication standard (OPC), etc.) for interfacing the PLCs~\cite{gupta2022end}. IGs are expected to provide computational resources for implementing Edge Computing strategies. It also requires to support of bidirectional communication channels to establish secure connections with 1) systems like SCADA system and MES, 2) Cloud-based applications and databases using protocols like MQTT, and REST and 3) network management systems (SNMP). Consequently, next-generation  IGs must fulfill non-functional requirements, such as they must be highly secure and reliable to prevent cyber attacks on the production processes from any external system and they can be easily configurable and scalable to overcome latency, redundancy, or load balancing issues.

In the I4.0 automation pyramid, \emph{Level 0}, i.e., field level, designates the production floor, which comprises machines, devices, actuators, and sensors. The field level is responsible for executing commands and collecting real-time production data that can be monitored at the higher levels. \emph{Level 1}, i.e., control level, manages sensors and actuators using a PLC or an equivalent one to make decisions for executing programmed tasks at the field level. Typically, field bus systems or Ethernet is used for connecting PLCs with field-level sensors and actuators. \emph{Level 2}, i.e., supervision level, employs the SCADA system for coordinating one or more PLCs at the control level. The SCADA environment also allows for storage capacities (e.g., SQL or NoSQL databases). It also provides interfaces that can be accessed remotely. 

\emph{Level 3}, i.e., planning level, utilizes manufacturing execution systems or manufacturing operation management systems. MES oversees the entire manufacturing process in a plant or factory, from raw materials to finished products including detailed production planning, production data acquisition, logistics, material management, quality management, safety processes inspection, and key progressive index determination. \emph{Level 4}, i.e., management level, uses ERP systems like SAP, Microsoft Dynamics, and {\color{black}Oracle Cloud Enterprise Resource Planning} for integrated management and production planning. ERP systems can be an extension of production planning systems. The management level is responsible for collective information about products, assets, customers, suppliers, contracts, offers, procurement, accounting, etc. 

{\color{black}
Thus, the integration of these levels can not only increase automation, efficiency, and production productivity in both manufacturing and industrial processes~\cite{sanchez2020autonomic} but also requisite for human-centered value addition, mass personalization, social fairness, and global demand for sustainable development leading I4.0 towards I5.0~\cite{aheleroff2022toward}.
}

\subsection{ARM TrustZone Technology}
ARM TrustZone technology aims at enabling a feature-rich open operating environment and a robust security solution within the System of Chip (SoC)~\cite{holdings2009arm}. Consequently, ARM TrustZone technology facilitates system-wide security by integrating protective measures into the ARM processor, bus fabric, and system peripheral IP using secure system architectures integrated for both the hardware and software components. 

\begin{figure}[!ht]
    \centering
    \includegraphics[width=.5\linewidth, keepaspectratio]{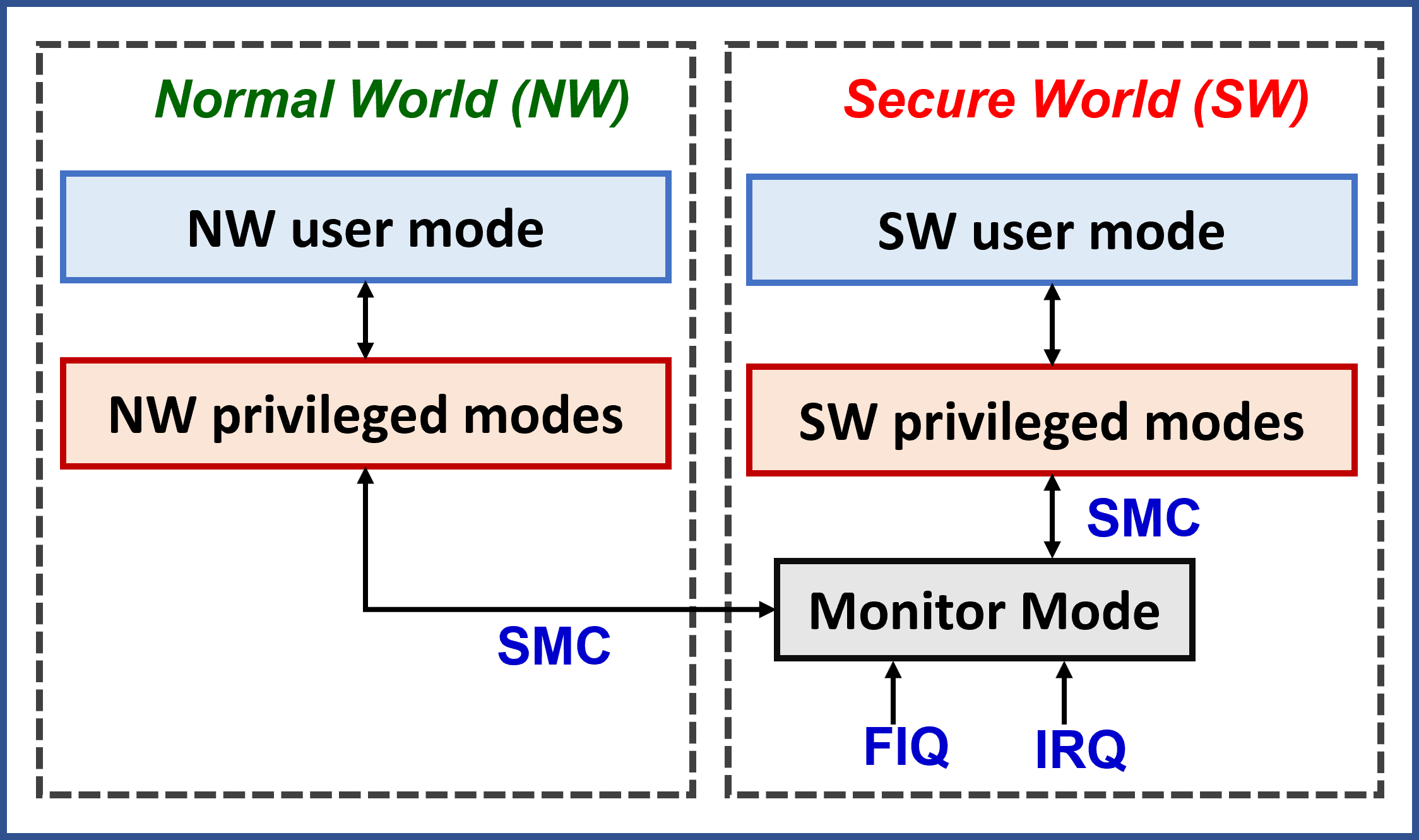}
    \caption{Modes in an ARM core implementing the security extensions (NW entry to monitor mode is tightly controlled. The entry to monitor can be triggered by SMC, IRQ, and FIQ to transfer execution to the SW)}\label{fig:TrustZone}
\end{figure}

Moreover, Trusted Execution Environments can provide key security mechanisms to ensure the confidentiality, integrity, and availability (CIA) of applications or services~\cite{cerdeira2020sok}. Figure~\ref{fig:TrustZone} depicts the TrustZone architecture that is designed using the principle of isolation providing two unconnected domains named Secure World (SW) and Normal World (NW). Cerdeira et al.~\cite{cerdeira2022rezone} explained that SW runs a smaller software stack consisting of Trusted Execution Environments (TEE) and Trusted Applications (TAs) and NW runs a rich software stack with full-blown OS and applications termed Rich Execution Environment (REE).

{\color{black}Typically, a physical processor core contains two virtual states, i.e., NW and SW, to serve `non-secure' and `secure' applications, respectively that can be interfaced using monitor mode~\cite{yang2020demystifying}. The role of the monitor is to provide a gatekeeper managing the switches between the NW and SW.} Monitor mode can either be triggered by software executing a dedicated Secure Monitor Call (SMC) instruction or hardware exception (interrupts (IRQ), fast interrupts (FIQ), and external abort) mechanisms.

\subsubsection{TrustZone Hardware Architecture}
A hardware architecture incorporates the security infrastructure throughout the system design based on the concept of a trusted platform. As a result, an end-to-end security solution that includes functional units and the debug infrastructure throughout the hardware design can be achieved.

Primarily, TrustZone Hardware Architecture uses the principle of separation, thus, partitioning all of SoC's hardware and software resources creates two worlds, i.e., the Secure world for the security subsystem, and the Normal world for everything else. Secondly, the extensions (e.g., Symmetric Multi-Processing mode, Asymmetric Multi-Processing mode) are implemented in some of the ARM processor cores that enable a single physical processor core to execute code from both the Normal world and the Secure world in a time-sliced fashion safely and efficiently. Lastly, security-aware debug infrastructure enables Secure world debug access without impacting debug visibility of the Normal world. The TrustZone debug extensions separate the debug access control, i.e., Secure privileged invasive debug, Secure privileged non-invasive (trace) debug, Secure user invasive debug, and Secure user non-invasive debug.

\subsubsection{TrustZone Hardware Library}
TrustZone Hardware Library has implemented system IPs that provide in-built support for the Security Extensions. AMBA3 Advanced eXtensible Interface (AXI) compliant bus matrix connects all of the system components for system-wide isolation. The AXI bus generator is complemented by a series of support components that include register slices for timing isolation, width scaling downsizers for reducing bus width to low bandwidth SoC regions, and synchronous or asynchronous bridges for linking clock domains.

\subsubsection{TrustZone Software Architecture}
In ARM architecture, Security Extensions are an open component that enables the customization of the Secure world software environment as per a given requirement. A secure world software stack on a TrustZone-enabled processor core can be implemented by using a Secure world operating system or by placing a synchronous library of code. 

A dedicated operating system can provide concurrent execution of multiple independent Secure world applications, run-time download of new security applications, and Secure world tasks that can be fully isolated from the Normal world environment. Synchronous library implementation enables single task execution at a time in the Secure world that can be entirely scheduled and managed using software calls from the Normal world operating system. However, the Secure world in these systems is a slave to the Normal world and cannot operate independently.

\section{Related Work}\label{sec:Relatedwork}
This section presents the relevant related work that focused on IGs' security aspects. Cheruvu et al.~\cite{cheruvu2020iot} investigated IoT framework gateways architectures from security perspectives. The authors stated that framework gateways must address two important security questions: 1) is the gateway trustworthy to connect IT and OT domains? and 2) how effective are authentication and authorization mechanism to protect CIA properties? Sauer et al.~\cite{sauer2022current} analyzed prevalent IIoT platforms for Industry 4.0 applications covering security, communication protocols, openness, extensibility, use of digital twins, device management, or cloud connectivity. The authors reported that Microsoft Azure IoT Suite leveraged TrustZone and Software Guard Extensions technologies.

Astarloa et al.~\cite{astarloa2016intelligent} proposed an intelligent-gateway IP implemented on an All-Programmable System-on-Chip (SoC) offering transparent operation between both worlds, i.e., machines and information technology. The solution addresses the difficulty of connecting heterogeneous systems, e.g., Ethernet ports with varying speeds to implement regular Ethernet or Industrial, serial ports like Modbus and Profibus. The potential security mechanism can be the authentication of users and devices using IEEE 802.IX combined with RADIUS, implementation of cryptographic libraries (such as OpenSSL) to secure protocols and applications for data interchange. Corradi et al.~\cite{corradi2021sirdam4} proposed a support infrastructure for reliable data acquisition and management in I4.0 that grants stakeholders safe and selective access to valuable data reflecting the production dynamics.

Bienhaus et al.~\cite{bienhaus2019gateway} proposed a security architecture for a gateway connecting the production level (e.g., sensors and actuators) and cloud systems. The security architecture exploited Trusted Platform Module (TPM) 2.0 providing secure storage and communication protocols (e.g., OPC UA or MQTT) within Transport Layer Security (TLS) using cryptographic keys. TPM can leverage password authentication or enhanced authorization policies to authorize cryptographic keys. Birnstill et al.~\cite{birnstill2017introducing} proposed integration of TPM-based cryptographic functions into the OPC unified architecture with a focus on remote attestation, i.e., establishing trust in a system's integrity. The authors described that dedicated hardware units providing security-related functions in TPM are hard to attack, and thus, can be integrated into OPC UA via ConformanceUnits or profiles. And, these profiles can be used by clients and servers for negotiating the parameters of a session.

Nguyen-Hoang~\cite{nguyen2019development} provided an open-source industrial IoT gateway framework that exploits Docker's micro-service to fill the IT/OT gap. The framework supports Siemens S7, Modbus RTU/TCP and IoT protocol MQTT (for IBM Watson), and REST API (for AWS EC2) industrial protocols implemented using NodeRED programming. Corotinschi and G{\u{a}}itan~\cite{corotinschi2018enabling} proposed an edge gateway for extending the connectivity of Modbus devices networks to an industrial network. The edge gateway exploited a microservice-based architecture to implement a generic object compiler, refreshing data module, cloud connectivity module, and database export utility for local Modbus data processing. kasinathan et al.~\cite{kasinathan2021secure} proposed a Workflow-Driven Security Framework (WDSF) that can guarantee CIA objectives for remote maintenance operations. However, TEE can be useful to enhance overall security by isolating the execution of security-sensitive transactions of the workflows at run-time. Table~\ref{tab:Comparison} compares this work with some of the recent related work in terms of security principles covered by them.

\begin{table}[!ht]
    \centering
    \footnotesize
    \caption{A comparison with related work in terms of security principles discussed}\label{tab:Comparison}
    \begin{tabularx}{1\linewidth}{p{.23\linewidth}p{.18\linewidth}p{.13\linewidth}p{.12\linewidth}p{.12\linewidth}}\hline
    \textbf{Reference} & \textbf{Secure execution environment} & \textbf{Boundary protection} & \textbf{Access control policies} & \textbf{Auditing and reviews}\\\hline
    This work & \checkmark & \checkmark & \checkmark & \checkmark \\\hline
    Astarloa et al.~\cite{astarloa2016intelligent} & - & - & \checkmark & \checkmark \\\hline
    Bienhaus et al.~\cite{bienhaus2019gateway} & \checkmark & - & - & -\\\hline
    Birnstill et al.~\cite{birnstill2017introducing} & \checkmark & - & \checkmark & -\\\hline
    Yu et al.~\cite{yu2021research} & - & \checkmark & - & - \\\hline
    kasinathan et al.~\cite{kasinathan2021secure} & - & - & \checkmark & \checkmark \\\hline
    Nguyen-Hoang~\cite{nguyen2019development} & - & - & - & - \\\hline
\end{tabularx}
\end{table}

\section{Pilot Study}\label{sec:PilotStudy}
In this section, we present a remote production-line maintenance\footnote{\url{https://www.collabs-project.eu/wp-content/uploads/2022/03/D5.2_COLLABS_Minimum_Viable_Product.pdf}} use case and enumerate security threats and vulnerabilities in it using the STRIDE threat modeling approach~\cite{shostack2014threat}. The motivation behind this study is to derive security principles that can be efficacious for protecting the proposed EC-IG against potential cyber-attacks.

\subsection{Remote Production-line Maintenance}\label{sec:RemoteMaintenance}
Timely maintenance of a production line (i.e., field level and control level) is critical for sustainable performance of production processes and the quality of finished products~\cite{sipsas2016collaborative}. Moreover, the external suppliers can also get controlled access to the industrial network which can minimize inter-dependency issues. Remote maintenance can be beneficial for optimizing maintenance decisions (i.e., minimizing unplanned downtime, operational hazards, and reducing disruptions to plant production) and incorporating predictive maintenance intelligence (i.e., identifying technical constraints in the manufacturing activities or assets-related issues optimizing asset lifespan and maintenance costs)~\cite{kasinathan2021secure}. Waidner and Kasper~\cite{waidner2016security} stated security in the industrial sector is also related to operational safety, e.g., protection of the physical environment including people, assets, and infrastructure from accidents, unintended mistakes, or lack of diligence. However, it is expected that IG must address the security concern mentioned below for remote maintenance of a production line. 
\begin{enumerate}[leftmargin=*]
    \item Guarantee confidentiality and integrity of proprietary information, sensitive data, or intellectual property.
    \item Maintain applications and data integrity against common cyber attacks. 
    \item Implement stringent data segregation schemes to prevent external access and transfer of confidential operational data.
    \item Use robust authentication and authorization mechanisms to prevent any illegitimate access.
\end{enumerate}

\subsection{Threat Analysis}
\begin{figure}[!ht]
    \centering
    \includegraphics[width=1\linewidth, keepaspectratio]{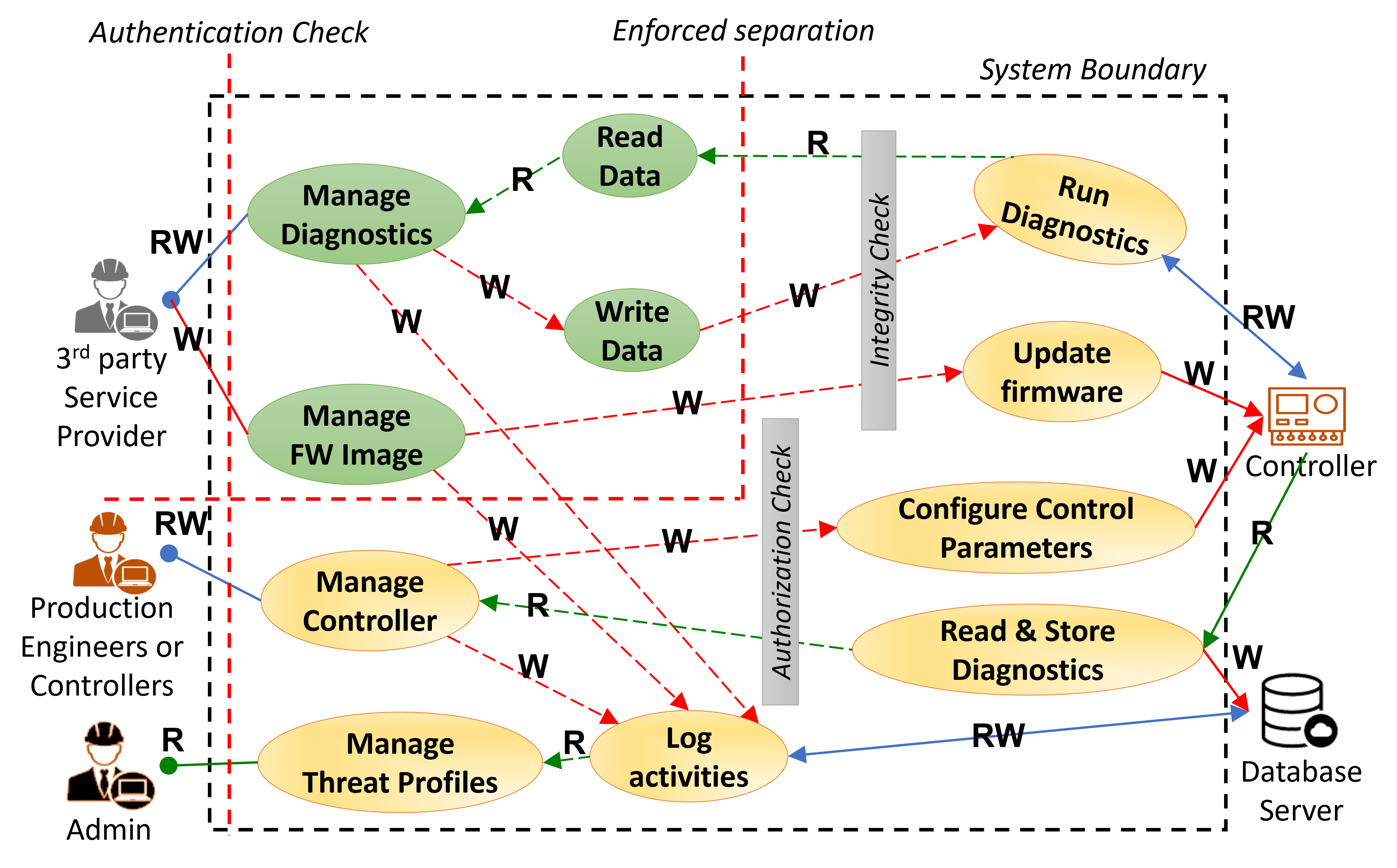}
    \caption{Remote production-line maintenance activities' data flow diagram (R = Read, W = Write, RW = Read and Write)}\label{fig:DataFlow}
\end{figure}

Figure~\ref{fig:DataFlow} illustrates the data flow diagram of the activities implemented in the prototype required for successful remote production-line maintenance. The data flow diagram of each activity essential for remote production-line maintenance can be highly useful for threat modeling and trust boundary identification~\cite{shostack2014threat}.
Legitimate users (actors) for remote production-line maintenance can be administrators, PLC controllers, production engineers, and third-party service providers. They can be responsible for managing controllers, running diagnostics, upgrading firmware images, or analyzing threat profiles, remotely. 

Third-party service providers are responsible for diagnostics management and are permitted to perform read-and-write operations of non-confidential data. Also, third-party service providers are required to upgrade PLC controllers' firmware at \emph{Level 2}. The production engineers or controllers are responsible for managing PLC controllers including the configuration of their running parameters and the activation of specific control tasks execution. The administrator is responsible for analyzing threat profiles.

However, the introduction of Internet technologies and multi-connectivity in IG exposes new attack surfaces~\cite{bienhaus2021secure}. Considering the high importance of production/manufacturing data, there can be several adversaries (e.g., black-hat hackers, industrial espionage) willing to circumvent the industrial gateway by exploiting existing vulnerabilities or architectural design flaws. Table~\ref{tab:ActivitiesMapping} lists legitimate users that can execute the supported activities and potential adversaries that can impact the confidentiality, integrity, and availability of EC-IG. Adversaries that can adversely impact a particular activity are assumed based on the activity's susceptibility and availability to fulfill a given requirement. Common security risks can be illegitimate access to sensitive information (e.g., spoofing, eavesdropping), intellectual property theft (e.g., espionage, malware),  production process manipulation or sabotage (e.g., ransomware, denial of service, unintended interactions between the sensitive assets and the third-party entities), and breach of communication interfaces connecting the gateway and the controllers (e.g., bypass firewalls).
\begin{table}[!ht]
    \centering
    \footnotesize
    \caption{Activities, legitimate users, and adversaries}\label{tab:ActivitiesMapping}
    \begin{tabularx}{.98\linewidth}{p{.28\linewidth} p{.27\linewidth} p{.27\linewidth}}\hline
    \textbf{Activities} & \textbf{Legitimate Users} & \textbf{Adversaries}\\\hline
    Diagnostic Management & Third-party service providers & Black-hat hackers, malware, malicious applications.\\\hline
    PLC controllers firmware upgradation & Third-party service providers & Malware, malicious applications.\\\hline
    PLC controllers management & PLC controllers, production engineers & Impostors, knowledgeable insider.\\\hline
    Administration, Threat profiles monitoring & Administrator & Impostors, knowledgeable insiders.\\\hline
    \end{tabularx}
\end{table}

We rely on the STRIDE threat modeling approach for analyzing threats to various remote production-line maintenance activities illustrated in Figure~\ref{fig:DataFlow}. Table~\ref{tab:ThreatModel} describes a correlation between the typical software elements, i.e., external entities, processes, data flows,  data stores, and the threats, i.e., spoofing, tampering, repudiation, information disclosure, denial of service, and elevation of privilege to maintain properties like authentication, integrity, non-repudiation, confidentiality, availability, and authorization, respectively~\cite{plaga2019securing}. Thus, the identification of potential threats leads to selecting design principles that can minimize the overall security risks.
\begin{table}[!ht]
    \centering
    \footnotesize
    \caption{Correlation between the STRIDE standard assets and threats}\label{tab:ThreatModel}
    \begin{tabularx}{.98\linewidth}{p{.45\linewidth}p{.02\linewidth}p{.02\linewidth}p{.02\linewidth}p{.03\linewidth}p{.03\linewidth}p{.03\linewidth}}\hline
    \textbf{Elements} & \rotatebox{90}{\textbf{Spoofing}} & \rotatebox{90}{\textbf{Tampering}} & \rotatebox{90}{\textbf{Repudiation}} & \rotatebox{90}{\parbox[t][][t]{1.2cm}{\textbf{Information Disclosure}}} & \rotatebox{90}{\parbox[t][][t]{1cm}{\textbf{Denial of Service}}} & \rotatebox{90}{\parbox[t][][t]{1.2cm}{\textbf{Elevation of Privilege}}}\\\hline
    External entities (activities or events responsible for managing processes) & \checkmark & & \checkmark & & &\\\hline
    Processes (applications or programs under execution) & \checkmark & \checkmark & \checkmark & \checkmark & \checkmark & \checkmark \\\hline
    Data flows (communication channels) & & \checkmark & & \checkmark & \checkmark & \\\hline
    Data stores (databases, caches, or transient storage) & & \checkmark & \checkmark & \checkmark & \checkmark &\\\hline
\end{tabularx}
\end{table}

\section{Design Principles, Edge Computing Requirements and Architecture}\label{sec:Architecture}
This section describes the design principles, Edge Computing (EC) requirements, and the architecture of the proposed EC-IG prototype.

\subsection{Design Principles}\label{sec:DesignPrinciples}
As the size and complexity of software systems increase, it is essential to incorporate strategies that can address potential threats and overcome inherent vulnerabilities effectively. The design principles that we apply for securing critical elements and minimizing the overall attack surfaces are explained below.

\begin{enumerate}[leftmargin=*]
    \item \textit{Secure execution environment}: Hardware-based trusted execution environment can provide a secure and isolated execution environment to prevent unauthorized access or modification of applications and data at run-time. Complete execution isolation for secure and non-secure applications and a secure channel for exchanging data between them can guarantee the confidentiality and integrity of data. The partition can also protect a secure application's embedded data (e.g., cryptographic keys) from being accessed by other applications. Moreover, features like remote attestation allow an administrator to remotely check the integrity of a trusted computing platform. Remote attestation can be a highly effective security solution for Edge computing systems to detect adversarial malware presence and verifies the state of EC-IG~\cite{gindre2021leveraging}.

    \item \textit{Access control mechanisms}: Access control mechanisms typically involve identification, authentication, and authorization by clearly specifying access to information or resources to a client only on a need basis. Authentication can maintain confidentiality and non-repudiation checks~\cite{gupta2020next}. Whereas, authorization ensures that only legitimate users can access specific information or resources to prevent tampering and information disclosure~\cite{sangchoolie2020analysis}.
    
     To implement access control mechanisms for managing the activities requested by trusted entities, the least privilege principle and least functionality principle are recommended. The least privilege principle limits the user privileges of an authorized person or processes to the minimum required level to perform specified activities. This can ensure that adversaries cannot take advantage in case the intrusion is successful. The least functionality principle ensures that the processes expose only required APIs (Application Programming Interfaces) and must prohibit or restrict non-essential functionalities and capabilities to prevent potential abuses or vulnerabilities exploitation.
     
    \item \textit{Boundary protection}: Boundary or perimeter protection can be a highly effective solution for segregating IT and OT. It can be useful to restrict malicious or unauthorized communications by defining a clear trust boundary for each legitimate client, i.e., third-party partners, PLC controllers, production engineers, and administrators. Boundary protection can be implemented using firewalls, encrypted tunnels, end-to-end secure channels, or gateways~\cite{stouffer2015nist}.
    
    Firewalls can monitor and control communications at the system boundary. Encrypted tunnels can ensure network channel integrity and prevent attacks like eavesdropping, or man-in-the-middle. An end-to-end secure channel is vital for providing access to end-users situated at separated locations and guaranteeing that data is not compromised. Lastly, a gateway can be a device supporting an inter-network connection by isolating IT and OT systems to fulfill the business objectives.
    
    \item \textit{Audits and reviews}: Audit trails can be useful for detecting security violations, intrusions, performance degradation, and system flaws, thereby improving the system's reliability and availability. Particularly, an audit trail can provide a specific location and trace information to determine security breaches or cyber-attacks, and it can improve non-repudiation by monitoring entities' actions and incidents.  
    
    The audit trail fundamentally relies on time-stamped sequential logs or traces generated by various processes running in a system. Therefore, persistent chronological logs and traces can assist significantly in forensic investigation or anomaly detection. Further logs can provide useful information for rectifying underlying vulnerabilities in the system. Consequently, we introduce a concept of an intelligent threat profiling mechanism that can be fed to intrusion detection systems for analyzing system usage, e.g., network traffic patterns, discrepancies, anomalies, etc., to detect possible attack scenarios.
\end{enumerate}

\subsection{Edge Computing Requirements}
Studies have reported that size, communication burden, and autonomous ability are fundamental principles of edge generation~\cite{zhang2020multilevel}. Size should be moderate and satisfy the condition $2 \leq Count_{feeder} \leq 10$ in a grid. Upstream traffic ($T_{up}$) of distribution services should not exceed the upstream bandwidth capacity of a wireless private network, i.e., $25Kbps \leq T_{up} \leq 186Kbps$. Autonomous ability includes controllable unit adjustment ability and self-healing ability after the malfunction inside the edge. Further, minimizing the response time, power consumption, and bandwidth usage must be considered when designing an EC-IG~\cite{chen2018edge}. Important features leveraged by the state-of-the-art edge frameworks for delivering localized computing and data storage are described below~\cite{atos2021edge}.

\begin{itemize}[leftmargin=*]
\item \textit{Support processing}: Provide operational capabilities as per a given use case to process (i.e., filter, compute, analyze) data and support machine learning models optimized for specific hardware. 
\item \textit{Communication}: IG must support bidirectional communication channels and standard industrial protocols for interfacing the programmable logic controllers (PLCs) and higher levels like SCADA, MES, and ERP. It also requires focusing on the security aspects of the supported communication protocols.
\item \textit{Modularity}: Modularization can be achieved by decoupling the core computing platform (i.e., heterogeneous acceleration platform), communication modules (i.e., wired or wireless network communication protocols and standards), and management features (i.e., remote attestation, autonomous ability).
\item \textit{Manageability}: IG is critical for horizontal and vertical integration for IOT4.0 to accelerate productivity and improve operational activities. Therefore, controllable unit adjustment ability, self-healing ability, and over-the-air updates are required for managing edge devices remotely.
\item \textit{Orchestration}: Orchestration simplifies management and enables distributed deployments resulting in negligible overhead to the system and minimal services downtime. 
\item \textit{Security}: Integrated security and privacy mechanisms, standards, platforms, and policies that can support the CIA properties. 
\end{itemize}

\subsection{Edge Computing based Industrial Gateway Prototype Architecture}
Architecture can be useful for drawing a blueprint of complex systems. High-level abstraction of complex systems at the functional and communication levels can assist security specialists or subject matter experts in analyzing potential threats and inherent vulnerabilities to employ strategies for addressing possible cyber attacks~\cite{gupta2019risk}. For a remote maintenance use case, EC-IG is required to facilitate bidirectional wired and wireless communications, data collection and storage, diagnostic and maintenance services, and threat profile creation.

\begin{figure}[!ht]
    \centering
    \includegraphics[width=.8\linewidth, keepaspectratio]{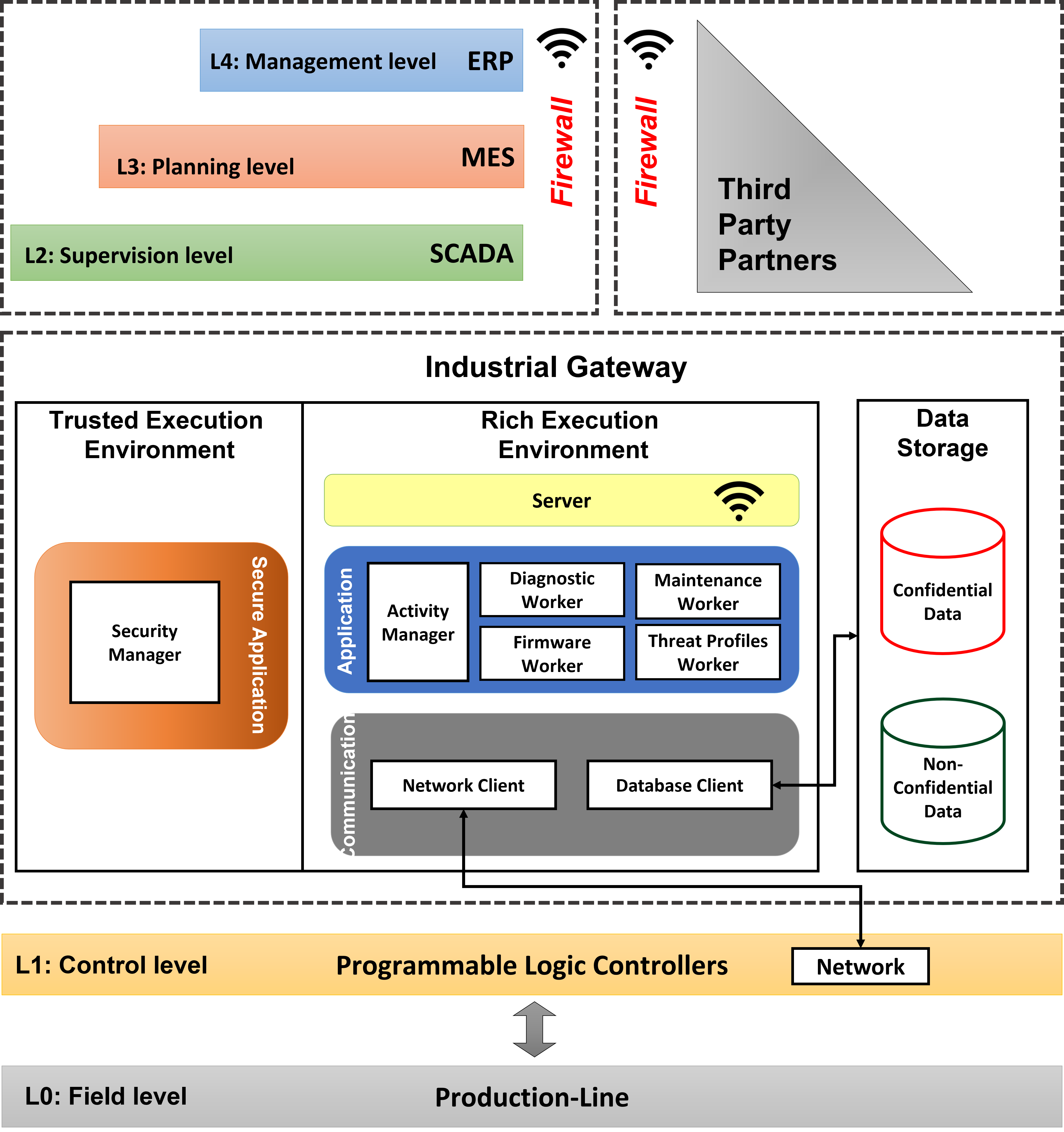}
    \caption{Edge computing based industrial gateway prototype architecture}\label{fig:RefArch}
\end{figure}

Figure~\ref{fig:RefArch} presents an architecture of the EC-IG prototype enabling horizontal and vertical integration between field- and control levels and supervision-, planning-, and management levels, and third-party partners. It can be observed that EC-IG consists of TEE, REE, and secure data storage. REE implements \textit{Server} (SVR), \textit{Activity Manager} (AM), \textit{Network Client} (NC) and \textit{Database Client} (DC). Whereas TEE is used for implementing security-sensitive applications, i.e., \textit{Security Manager} (SM), that must not be directly exposed to external entities.

TEE exploits physical separation on hardware devices to create a secure partition. Interaction between applications residing in REE and applications residing in TEE is managed using strict access control policies that are also implemented at the hardware level to segregate both execution environments. Eventually, TEE can provide (1) isolated execution, (2) a trusted path, (3) secure storage, (4) secure provisioning, and (5) local and remote attestation~\cite{ekberg2014untapped}.

Individual responsibilities of SVR, AM, NC, DC, and SM are described below.
\begin{itemize}[leftmargin=*]
    \item SVR exposes REST APIs for providing access to clients. Under REST architecture, all interactions are initiated by the client, i.e., the server can respond to clients only after receiving their requests. Also, SVR is responsible for executing some periodical activities described in Sections~\ref{sec:ControlManagement} and~\ref{sec:ThreatProfiles}. 
    
    \item AM is implemented in NW. AM encapsulates \textit{Command Parser} (CP), \textit{Diagnostic Worker} (DW), \textit{Firmware Worker} (FM), \textit{Maintenance Worker} (MW), \textit{Threat Profiles Worker} (TPW) to serve the incoming request from the verified clients forwarded by SVR.
    \item NC supports different protocols like Modbus or Ethernet connecting EC-IGs with PLCs. NC provides simple APIs like connect, disconnect, read and write supporting RTU (serial) and TCP (Ethernet) communications. 
    \item DC enables read and write operations to and from local storage. DC encrypts and decrypts the data during every write and read operation to the local storage.
    \item SM is implemented in SW. SM verifies AM's signature and validates the addresses requested by third-party partners to perform diagnostic operations. SM also validates the keys received from SVR to execute periodical commands and from production engineers and controllers to perform control-level maintenance.  
\end{itemize}

\section{Prototype Design and Implementation}\label{sec:Prototype}
This section presents the high-level design of the EC-IG prototype delineating the main features for remote production-line maintenance as a proof-of-concept and the prototype implementation details on the OP-TEE-enabled QEMU platform. 

\subsection{High-level Design Details}
Table~\ref{tab:ActivitiesDetails} lists the main activities supported by the prototype and the commands that authenticated users can use to perform the remote maintenance activities.
\begin{table}[!ht]
    \centering
    \footnotesize
    \caption{Activities supported}\label{tab:ActivitiesDetails}
    \begin{tabularx}{1\linewidth}{p{.25\linewidth} p{.44\linewidth} p{.18\linewidth}}\hline
     \textbf{Activities} & \textbf{Command Example} & \textbf{Users}\\\hline
    Remote diagnostic management restricted to non-confidential data & \texttt{read $<$asset id$>$ $<$addr$>$ $<$length$>$; write $<$asset id$>$ $<$addr$>$ $<$length$>$ $<$space separated hex data$>$}. & Third-party service providers\\\hline
    Remote firmware management & \texttt{update $<$asset id$>$ $<$filename$>$}. & Third-party service providers\\\hline
    Remote control-level management & \texttt{read\_s $<$key$>$ $<$asset id$>$ $<$addr$>$ $<$length$>$; write\_s $<$key$>$ $<$asset id$>$ $<$addr$>$ $<$length$>$ $<$space separated hex data$>$; store\_s $<$key$>$ $<$asset id$>$}. & Production engineers and controllers\\\hline
    Threat Profiles Management & \texttt{gen\_threat\_profile\_s $<$key$>$}. & Administrator\\\hline
    \end{tabularx}
\end{table}

\subsubsection{Diagnostic Management}
\begin{figure}[!ht]
    \centering
    \includegraphics[width=1\linewidth, keepaspectratio]{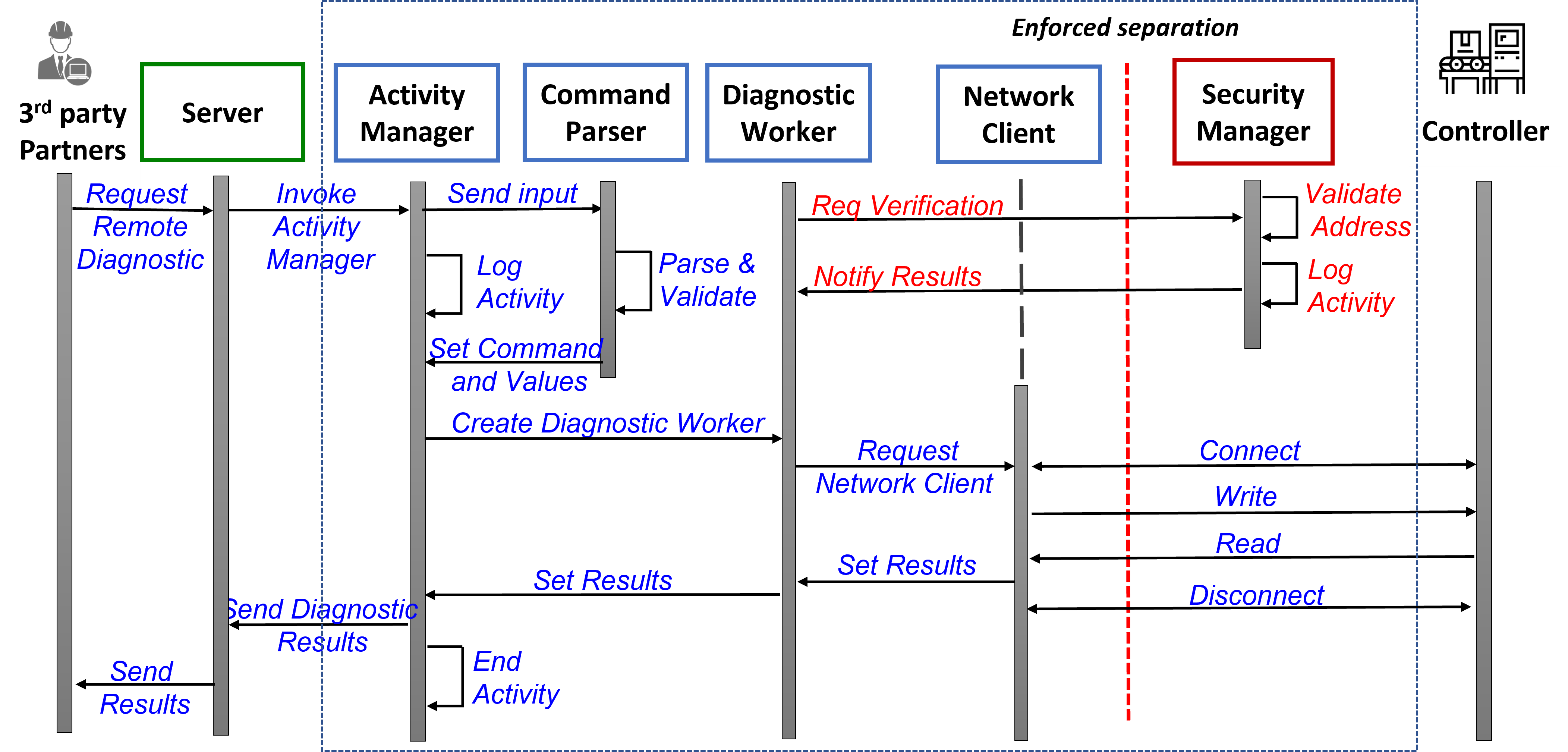}
    \caption{Sequence diagram for diagnostic management activity}\label{fig:ActDiag}
\end{figure}

Diagnostic management enables third-party service providers to perform remote maintenance activity at regular intervals. As explained in Section~\ref{sec:RemoteMaintenance}, remote diagnostics for industrial controllers (e.g., PLCs) maintenance is a critical requirement and inadequate maintenance can adversely impact both the performance of the production processes and the quality of the final products. Figure~\ref{fig:ActDiag} presents the sequence diagram for the remote maintenance activity describing the message exchange and interactions between the SVR, AM, CP, DW, NC, and SM to execute \texttt{read} and \texttt{write} commands. 

SVR first authenticates the third-party service providers to accept their incoming requests. After a successful user verification, SVR invokes AM's instance to execute the legitimate user's request. AM interacts with CM to parse and validate the input arguments into a well-defined structure. CM is also responsible for integrity checks to detect any inconsistency or inaccuracy in the input values. Subsequently, AM assigns the validated command to DW that internally requests SM residing in the TEE to validate the addresses for the read and write operations. Address validation is highly critical for preventing access to confidential data. DW instructs NC to interact with industrial controllers that involve connect, read/write, and disconnect steps, after the successful address validation. Finally, SVR notifies the activity results to the requester. Both AM and SM generate chronicle logs of the entire activity that is necessary for audit trails and reviews. 

\subsubsection{Firmware Management}
\begin{figure}[!ht]
    \centering
    \includegraphics[width=.8\linewidth, keepaspectratio]{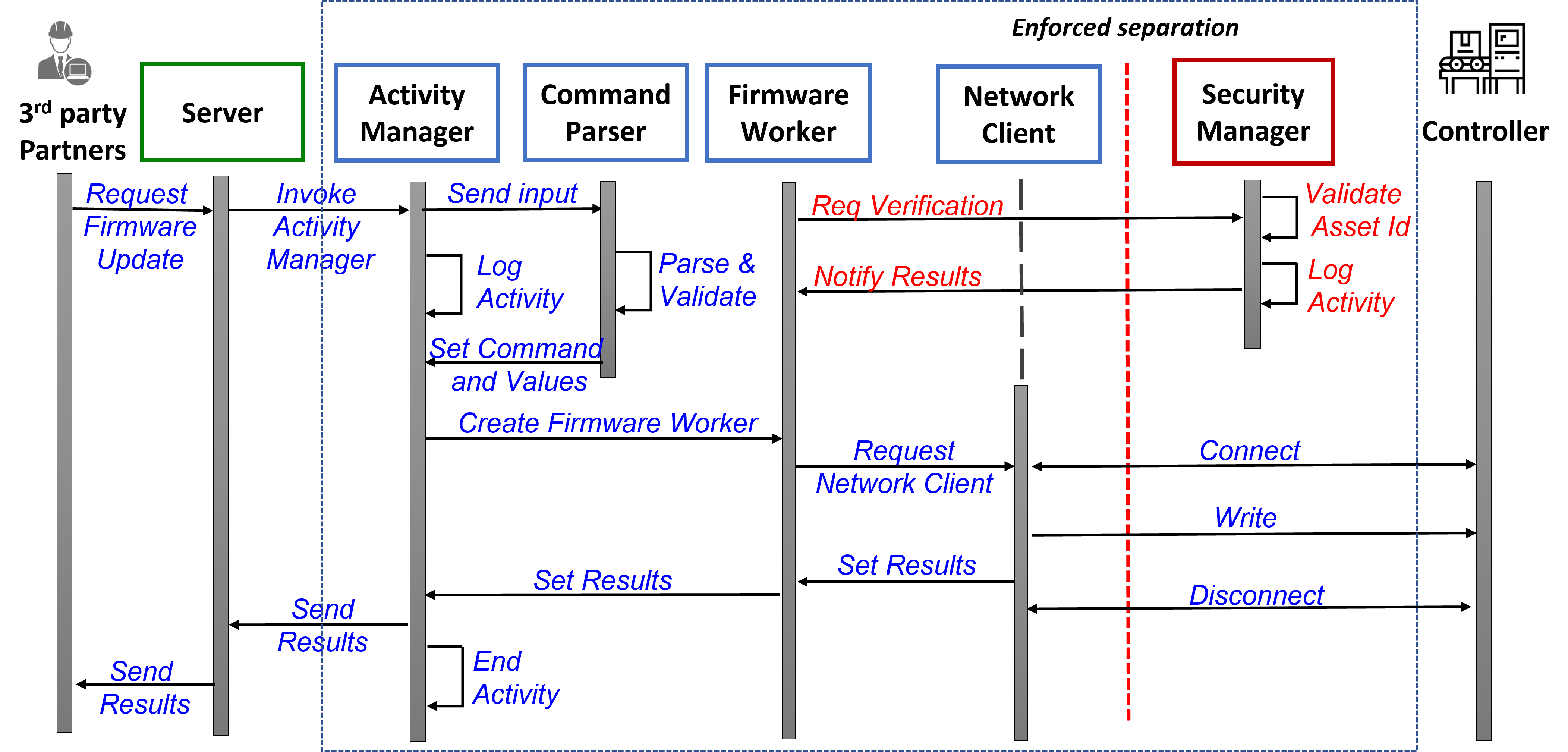}
    \caption{Sequence diagram for remote firmware update activity}\label{fig:ActFirmware}
\end{figure}

Firmware is one of the most trusted components for an industrial controller~\cite{cheruvu2020iot}. It is necessary that EC-IG must assert the firmware update requirement for a specified controller and must obtain verifiable proof of successful installation from that controller. Figure~\ref{fig:ActFirmware} presents the sequence diagram for the remote firmware update activity. SVR establishes the legitimacy of a user (e.g., Original Equipment Manufacturer) before invoking AM for firmware updates targeted for the specified industrial controller.  

AM requests CM to parse and validate the \texttt{update} command. After the successful command validation, AM instantiates FM to execute the given task. FM requests SM to validate the asset id before initiating the firmware transfer process by engaging the NC. Ultimately, AM obtains the verifiable proof of successful firmware update and sends the notification to SVR that propagates the installation proof further to the user.  

\subsubsection{Control-level Management}\label{sec:ControlManagement}
\begin{figure}[!ht]
    \centering
    \includegraphics[width=.8\linewidth, keepaspectratio]{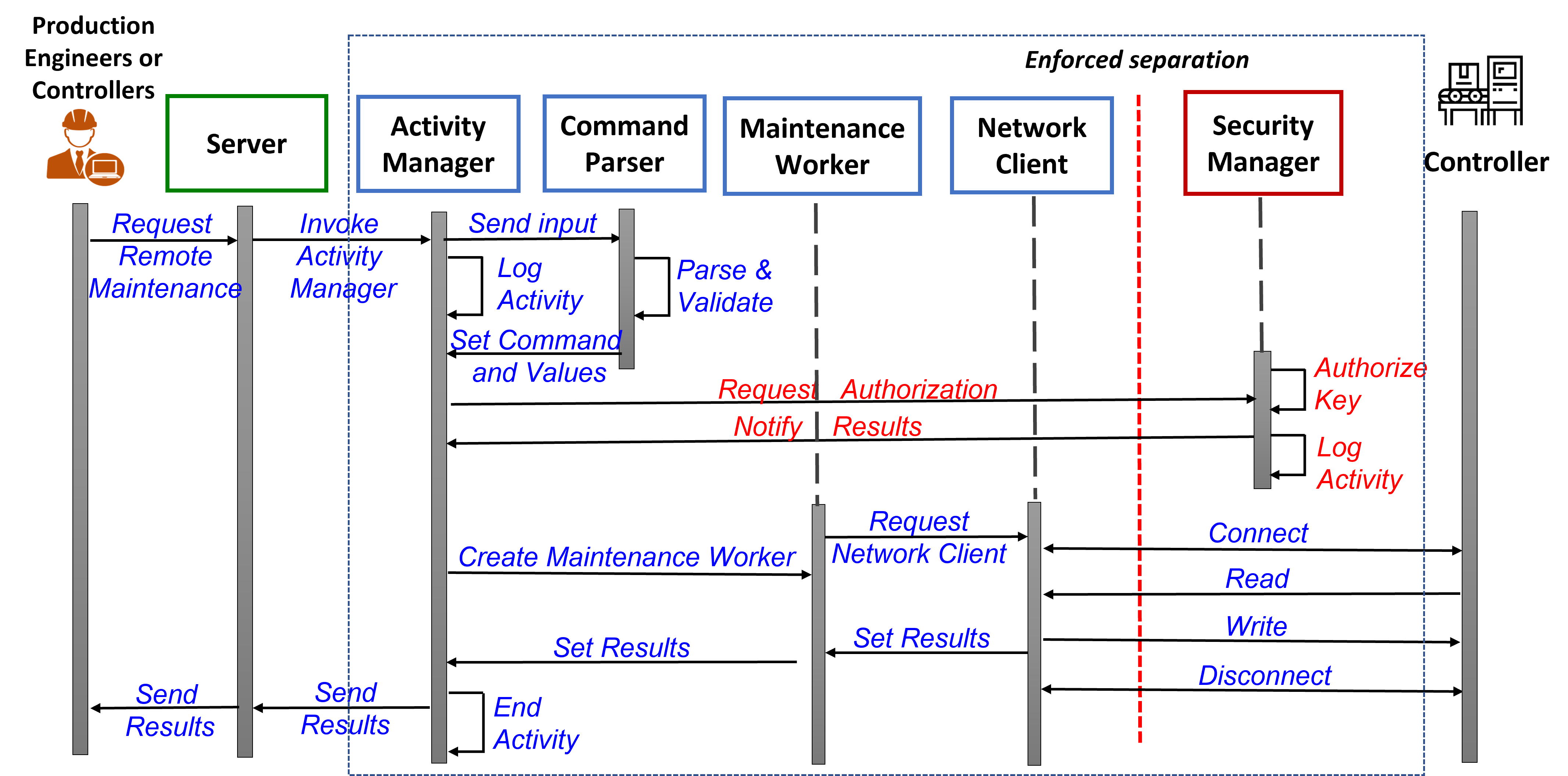}
    \caption{Sequence diagram for controller management activity}\label{fig:ActMaint}
\end{figure}

Control-level management, i.e., remote monitoring and predictive maintenance of industrial controllers at a control-level can not only synergize the control tasks but can also reduce the overall maintenance downtime. Moica et al.~\cite{moica2018change} described that envision, enact, and enable can be the essential elements for smart management of control- and field-level, which can be achieved with remote access to the control-level. Figure~\ref{fig:ActMaint} presents the sequence diagram for remote control-level management that explains how production engineers or controllers can configure control parameters and activate or deactivates control tasks using \texttt{read\_s} and \texttt{write\_s} commands.

SVR authenticates the incoming requests from production engineers and controllers to invoke AM that in turn requests CM to parse and validate the command. Since production engineers or controllers can access both confidential and non-confidential data, AM performs requests SM to perform an authorization check before instantiating MW. After a successful authorization check, MW instructs NC to perform the read/write operations on an industrial controller.

\begin{figure}[!ht]
    \centering
    \includegraphics[width=.8\linewidth, keepaspectratio]{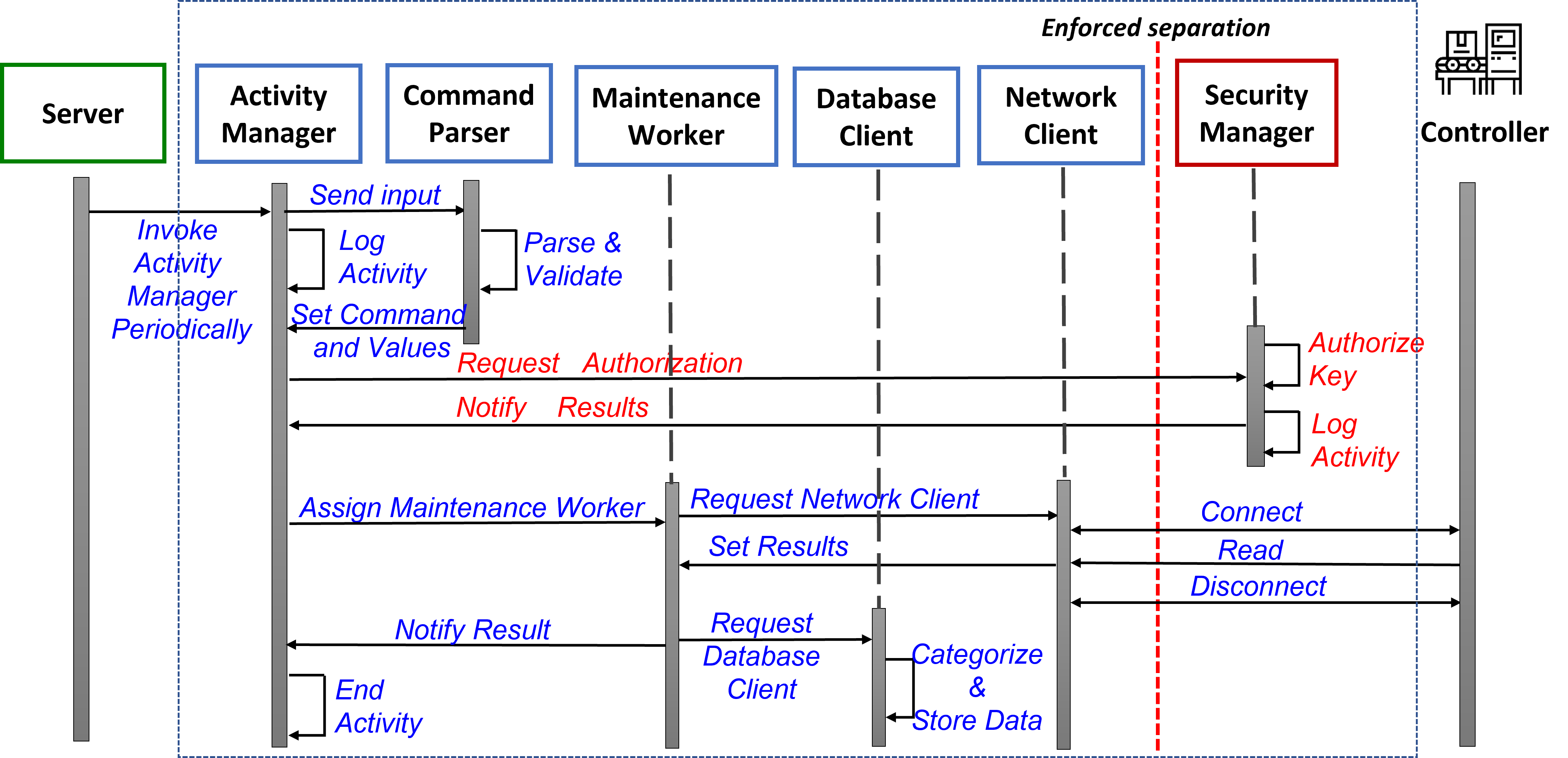}
    \caption{Sequence diagram for control-level record storing activity}\label{fig:ActRecord}
\end{figure}

Also, the prototype gives flexibility to production engineers and controllers to access control-level records that are periodically collected. Figure~\ref{fig:ActRecord} presents the sequence diagram for control-level record-storing activity. SVR periodically invokes AM to read data from controllers using \texttt{store\_s}. However, AM authorizes every request before engaging MW to read the data from controllers using NC.

DC employs the best available encryption algorithms for storing controllers' data locally. DC is also responsible for the categorization of confidential and non-confidential data that are stored separately in local storage. After successful local data storage, DC sends the results to AM which notifies SVR to send the encrypted data to authorized users using secure transfer. Moreover, production engineers and controllers can retrieve these records using a wired interface by physically accessing the EC-IG.

\subsubsection{Threat Profiles Management}\label{sec:ThreatProfiles}
\begin{figure}[!ht]
    \centering
    \includegraphics[width=.8\linewidth, keepaspectratio]{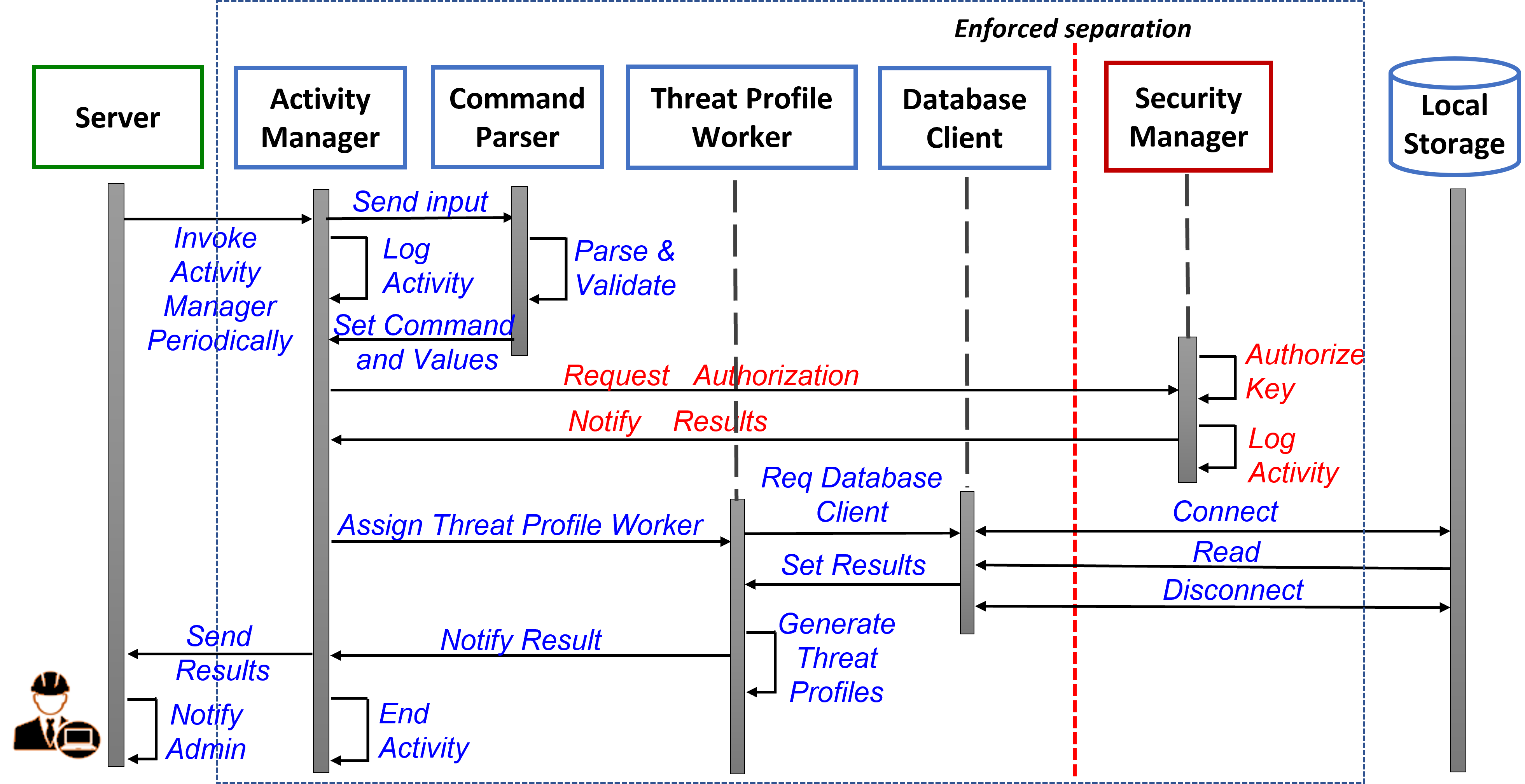}
    \caption{Sequence diagram for threat profiles generation activity}\label{fig:ActThreat}
\end{figure}

Threat profiling exploits the raw data generated by processes running on EC-IG that can be useful for administrators or can be fed to more specific applications for monitoring and analyzing uncertainties (e.g., denial-of-service, black-box decisions, local discrepancy), and common security problems (e.g., social-engineering attacks, insider attacks, and sensitive data leakage) to EC-IG.  Figure~\ref{fig:ActThreat} presents the sequence diagram for threat profile generation activity.

SVR periodically invokes AM to generate threat profiles using \texttt{gen\_threat\_profile\_s} command. AM requests SM to verify the key after CM successfully parses and validates the command. Subsequently, AM instantiates TPW that requests DC to get the locally stored log files for a requested period. TPW uses machine learning algorithms and data structures to generate threat profiles from the log files. As AM notifies the results to SVR, the report is sent to the administrator. 

\subsection{Implementation Details}
\begin{figure}[!ht]
    \centering
    \includegraphics[width=.6\linewidth, keepaspectratio]{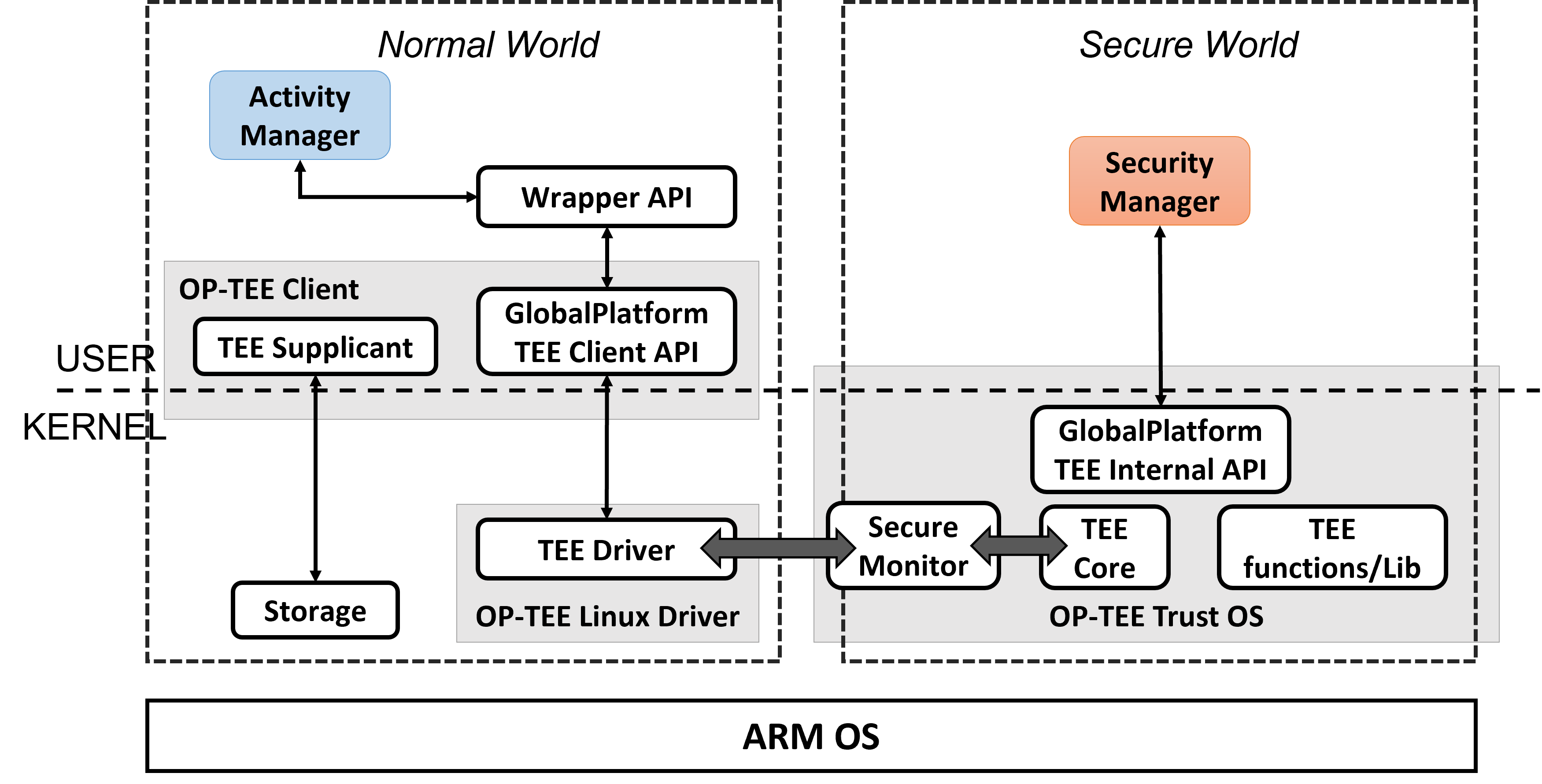}
    \caption{OP-TEE environment~\cite{yang2020demystifying}}\label{fig:OP_TEE_env}
\end{figure}

We rely on OP-TEE\footnote{\url{https://optee.readthedocs.io/en/latest/building/devices/qemu.html}} (Open Portable Trusted Execution Environment) which is an open-source project for developing the EC-IG prototype. OP-TEE is an open-source TEE that supports TrustZone technology on both Armv7-A and Armv8-A architecture. Figure~\ref{fig:OP_TEE_env} illustrates the OP-TEE environment consisting of three components, i.e., OP-TEE Client, OP-TEE Linux driver, and OP-TEE Trusted OS. 

Trusted OS implemented TEE's system services as C-based libraries and exposed C-based service APIs to the application layer. TEE Client and internal APIs for cross-world communication between NW applications and SW applications are listed in Table~\ref{tab:TEEAPIs}. The interaction between NW and SW applications involves four main data structures that are \textit{Context}, \textit{Session}, \textit{Command}, and \textit{Parameter}~\cite{wan2020rustee}. AM in REE (i.e., Linux) uses \textit{TEEC\_InitializeContext} to create logical connections that are called ``Context'' without requesting any specific trusted application to collaborate with the help of \textit{TA\_CreateEntryPoint}. In the next step, AM uses \textit{TEEC\_OpenSession} to create a logical container that is called ``Session'' for linking AM with SM in TEE (i.e., OP-TEE) by calling \textit{TA\_OpenSessionEntryPoint} and this Session is only valid under a registered Context.
\begin{table}[!ht]
    \centering
    \footnotesize
    \caption{TEE client and internal APIs}\label{tab:TEEAPIs}
    \begin{tabularx}{.9\linewidth}{|C|C|}\hline
    \textbf{TEE Client API} & \textbf{TEE Internal API} \\\hline
    \textit{TEEC\_InitializeContext} & \textit{TA\_CreateEntryPoint}\\\hline
    \textit{TEEC\_OpenSession} & \textit{TA\_OpenSessionEntryPoint}\\\hline
    \textit{TEEC\_InvokeCommand} & \textit{TA\_InvokeCommandEntryPoint}\\\hline
    \textit{TEEC\_CloseSession} & \textit{TA\_CloseSessionEntryPoint}\\\hline
    \textit{TEEC\_FinalizeContext} & \textit{TA\_DestroyEntryPoint}\\\hline
    \end{tabularx}
\end{table}

\begin{figure}[!ht]
    \centering
    \includegraphics[width=.6\linewidth, keepaspectratio]{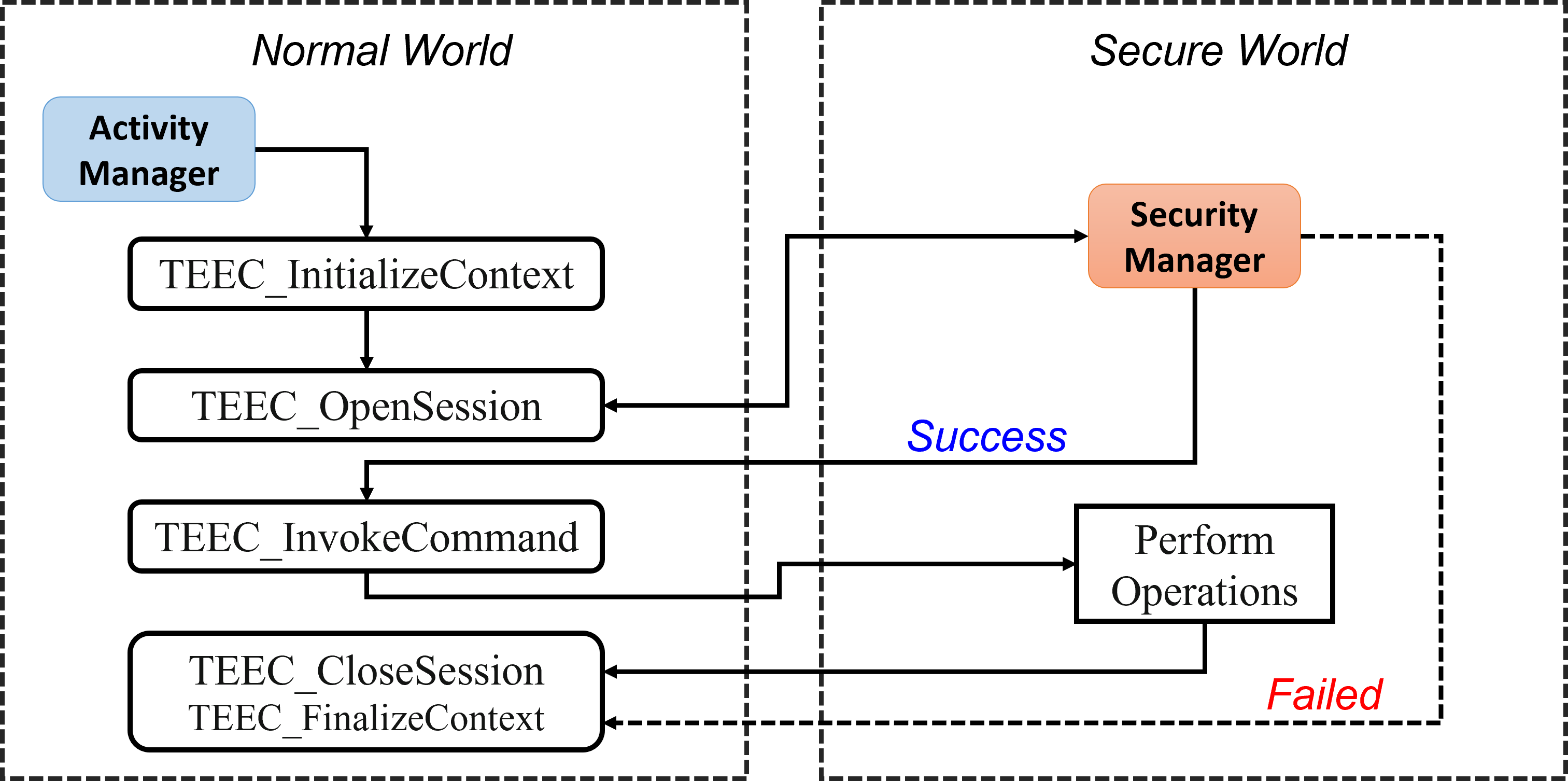}
    \caption{Verification process between AM and SM}\label{fig:AM_SM}
\end{figure}

Figure~\ref{fig:AM_SM} illustrates an interaction between AM and SM combined with the verification process for determining the legitimacy of AM to prevent any malicious access~\cite{jiang2017effective}. The verification process applies two independent operations, i.e. training and normal. During the training, the hash value of AM image registered as a shared file by the TrustZone driver is calculated using the SHA-1 algorithm. Subsequently, the calculated hash value of AM image is stored in TEE that is specified as \texttt{hash\_correct}.

During the normal operation, the verification process calculates the AM hash using the same SHA-1 algorithm once a session is established between AM and TM, which is specified as \texttt{hash\_calculated}. Both the \texttt{hash\_correct} and \texttt{hash\_calculated} are compared to verify the legitimacy of AM. If the verification process fails, the session is closed to protect the confidentiality and integrity of EC-IG. The successful authentication continues the session between AM and SM to perform the specified operation using call Commands, i.e., \textit{TEEC\_InvokeCommand} and \textit{TA\_InvokeCommandEntryPoint}. A Command can pass a maximum of four pairs of Parameters for using cross-world shared memory, i.e., sharing the plaintext/ciphertext across NW and SW. \textit{TEEC\_CloseSession} and \textit{TA\_CloseSessionEntryPoint} calls close the Session. Lastly, \textit{TEEC\_FinalizeContext} and \textit{TA\_DestroyEntryPoint} calls finalized the context, thus, concluding the communication between AM and SM.

\section{Discussions}\label{sec:Discussion}
Introducing Internet technologies in I4.0 is clearly a major technological advantage, however, it drastically increases the attack surfaces that are constantly exploited by adversaries. Thus, it is recommended to devise security solutions according to the adversaries' motivation and capabilities depicted in Figure~\ref{fig:adversaries} to achieve a desired security assurance level described by IEC 62443 standard~\cite{international2019iec}.

\subsection{TrustZone-assisted Systems}
In recent years, hardware-assisted isolated environments have evolved rapidly for designing security-sensitive systems including industrial control systems, servers, and low-end devices. TrustZone provides several security primitives that developers can leverage to implement trustworthy systems~\cite{pinto2019demystifying}. Gentilal et al.~\cite{gentilal2017trustzone} proposed a bitcoin wallet application that makes direct use of a TrustZone-enabled processor. The authors reported that the use of TrustZone technology makes the wallet resilient against dictionary and side-channel attacks, however, the performance of the read and write operations is slightly downgraded due to the use of encrypted storage. Zhang et al.~\cite{zhang2016case} designed a cache-assisted secure execution framework (CASE) that relies on a TrustZone-enabled SoC-Bound execution environment to protect against both cold boots and compromised rich OS attacks. The authors described that CASE can be operated in two modes, i.e., SoC-Bound execution in the normal cache and SoC-Bound execution in the secure cache, to balance the system's performance and security.

However, studies have identified successful attacks on TrustZone-assisted systems across various platforms. Cerdeira et al.~\cite{cerdeira2020sok} analyzed two hundred-seven TEE bug reports between 2013 and 2018 and discussed countermeasures that can be investigated to address architectural, implementation, or hardware issues. Koutroumpouchos et al.~\cite{koutroumpouchos2021building} classified the discovered vulnerabilities and related attacks into three main
categories, i.e., software, architectural, and side-channel attacks. The authors also provided generic attack paths illustrating possible vectors that can be exploited by adversaries to breach TrustZone security. Table~\ref{tab:AttackAnalysis} highlights vulnerabilities and risks that could arise due to incorrect OP-TEE usage at the architectural, implementation, and hardware levels.

\begin{table*}[!ht]
    \centering
    \caption{Vulnerabilities and risks}\label{tab:AttackAnalysis}
    \scriptsize
    \begin{tabularx}{1\linewidth}{p{.08\linewidth} p{.5\linewidth} p{.34\linewidth}}\\\hline
    \textbf{Category} & \textbf{Vulnerabilities} & \textbf{Risks}\\\hline
    Architectural & Tee Attack Surface, i.e., SW drivers run in the TEE kernel space, Interfaces between TEE system subcomponents, A large TEE trusted computing buses (TCBs). & SMC interface exploited by NW software access to TAs.\\\hline
    Architectural & Isolation between Normal and Secure Worlds, i.e., TAs can map physical memory in the NW, and Information leaks to NW through debugging channels. & Escalating privileges (by exploiting applications having privileged access to SW). This can also lead to fuzzing and MITM.\\\hline
    Architectural & Memory protection mechanisms, i.e., Absence or weak ASLR (Address Space Layout Randomization) implementation, No stack cookies, guard pages, or execution protection, Access to TEE-restricted memory. & Illegitimate root privileges in Linux kernel. Microarchitectural side-channels.\\\hline
    Architectural & Trust Bootstrapping, i.e., Lack of software-independent TEE integrity reporting, Ill-supported TA revocation. & Remote attestation, data sealing process, and trusted boot process can be affected.\\\hline
    Architectural & Hardware exceptions (SMC, IRQ, FIQ), caches, power management modules & Microarchitectural attack vectors.\\\hline
    Architectural & Semantic gaps when passing data between the TEE and the untrusted OS & Non-secure application exploits a trusted application to access a portion of memory that it does not own.\\\hline
    Implementation & Validation bugs (i.e., improper handling of input and/or output values) while implementing secure monitor, TAs, trusted kernel, and secure boot loader. & Unpredictable behavior or system crash.\\\hline
    Implementation & Functional Bugs (i.e., inconsistencies between a given specification and the implementation) resulting in buffer overflows, inadequate memory protection, improper configuration of peripherals, incorrect usage of security protocols, or cryptography primitives. & Affect the core functionality or result in unintended behavior.\\\hline
    Implementation & Extrinsic bugs resulting in concurrency issues, race conditions, or software side-channels. & Unpredictable outcomes.\\\hline
    Implementation & Improper interrupt prioritization handling in non-secure software & Denial-of-service (DoS) attacks\\\hline
    Implementation & Full disk encryption mechanism & Secure world privileged code exploitation.\\\hline
    Implementation & Bounds-checking SMC requests flaws (e.g., to cause Secure Execution Environment (SEE) to write controlled data to an arbitrary secure memory location) & kernel-level privileges exploitation.\\\hline
    Implementation & Self-modifying code and mismatched cacheability attributes & Cache-based attack vectors.\\\hline
    Hardware & Incorrect implementation of hardware Reconfigurable hardware (e.g., FPGA), Energy management mechanisms & Architectural Implications like Side-channels, exploiting the computation of SW operations, and secure boot program.\\\hline
    Hardware & Exploitation of caches, Branch predictor, e.g., branch target buffer (BTB) unit, Rowhammer, i.e., software-induced hardware fault that affects DRAM memories and enables an attacker to flip bits in physical memory by solely performing memory read operations. & Microarchitectural Side-Channels.\\\hline
    Hardware & DRAM and Memory protection & Physical attacks, e.g., memory or bus probing (MP)\\\hline
    Hardware & Recover cryptographic keys from local storage & Compromised local OS.\\\hline
    Hardware & Inadequate protection of operating limits of processors & Dynamic Voltage and Frequency Scaling for inducing faults.\\\hline
    \end{tabularx}
\end{table*}

Large-size TCBs (i.e., OS kernel, privileged services, and libraries) systems can be more vulnerable. Pinto and Santos~\cite{pinto2019demystifying} described a larger number of lines of source code and more complex inter-component interactions may inevitably inherit potential security deficiencies. Eventually, TrustZone provides an isolated environment to implement security-sensitive features that can be executed securely without the interference of the local (untrusted) OS. Also, the Platform Partition Controller must be (re)configured exclusively by a trusted component that can guarantee the enforcement of the memory access permissions~\cite{cerdeira2022rezone}. It is recommended that the access privileges to the trusted OS should be minimal to prevent unauthorized access or penetration attacks including sensitive data access, business operations disruption, and back-doors entry. 

\subsection{Access Control Mechanisms}
IEC 62443-4-2 standards specified that a client must be authentication and the intended remote maintenance activity requited by clients must be pre-authorized~\cite{international2019iec}. Robust authentication and authorization mechanisms can mitigate spoofing attacks as well as control unauthorized requests from improper entities, thus, improving EG-IG's availability and reliability~\cite{sangchoolie2020analysis}. According to the NIST Guide to Industrial Control Systems (ICS) security~\cite{stouffer2015nist}, whitelisting, i.e., granting access to known clients, instead of blacklisting, i.e, denying access to unknown clients, can be more logical and efficient to establish communications between authenticated and authorized source-destination pairs.

A distributed or centralized approach can be employed to authenticate and authorize a client. In distributed authentication and authorization, each system implements a local client verification process and grants privileges to resources based on predefined rules or policies. In centralized authentication and authorization, a centralized system is responsible for clients' authentication and an authentication protocol is used for establishing a secure channel between the source and destination. Strong authentication can minimize man-in-the-middle attacks

Salonikias et al.~\cite{salonikias2019access} investigated access control requirements in IIoT focusing on the diversity involved in the industrial environments. After that, the authors proposed an access control architecture capable of achieving access control using a layered approach based on virtualization concepts. Xue et al.~\cite{xue2019secure} proposed a lightweight and privacy-preserving authentication protocol between clients and edge devices using a combination of group signature and hash chain techniques. Similarly, Li et al.~\cite{li2014live} proposed a lightweight integrity verification and content access control mechanism that adopts one-way hash functions to produce content signatures such that integrity verification is done by verifying hash-based signatures. 

Ren et al.~\cite{xue2019secure} exploited blockchain technology to design identity management and access control mechanism to ensure the data security of IIoT. The authors further explained that attribute-based and role-based access control are not suitable for distributed IoT environments due to a lack of dynamic scalability. Marra et al.~\cite{marra2019distributed} presented a distributed usage control framework designed for distributed Peer-to-Peer (P2P) systems, without a root of trust, targeted for IIoT. Access policies for each activity to interact with the EC-IG must clearly specify the address range and read/write permissions.

\subsection{Boundary Protection Devices}
Boundary protection devices can be a vital component for IIoT security. Kayan et al.~\cite{kayan2021cybersecurity} described integration of IT to OT caused new vulnerabilities that are mainly due to weak boundary protection. Weak boundaries between OT and IT (enterprise) networks can be exploited by unauthorized access. The authors reported that the TRITON and German Steel Mill attacks are some examples of cyber-attacks that happened due to weak boundary protection.
Mosteiro-Sanchez et al.~\cite{mosteiro2020securing} recommended using a combination of Demilitarized Zone (DMZ) and firewalls to restrict external access to gateways isolating OT and IT networks. The authors described that the Next-Generation Firewalls (NGFWs) can offer application-level inspection providing greater control over what enters and leaves the OT network through IG. 

Yu et al.~\cite{yu2021research} proposed a mimic defense model for detecting security issues of boundary protection mechanisms. The model consists of an input and output module, a dispatch module, an arbitration module, and an awareness and control module responsible for different security measures. The input and output module manages direct communication with the networks for receiving and forwarding data packets and maintaining a buffer queue to mitigate DoS attacks. The dispatch module dispatches the pool of heterogeneous executors and schedules the switching, cleaning, and recovery of the executor pool by changing data distribution flow and online/offline instructions  The arbitration module is responsible for receiving the primary output from each executor and then establishing an isolated communication channel with each executor. It reports each final result to the awareness and control module.  The awareness and control module analyzes the results obtained from the arbitration module to initiate necessary actions as per the security policies employed.

\subsection{Threat Profiling Mechanism}
EC-IGs interfacing IT and OT are unarguably critical devices, thus, auditing and reviewing mechanisms can be useful to determine any abnormal behaviors, violations, or vulnerabilities in them. However, considering edge devices can have only limited computation and storage capabilities, a huge number of log generations for continuous monitoring can degrade the overall performance. Furthermore, offline log diagnostics can be time-consuming and error-prone due to their inherent complexity and distributed nature. Therefore, a real-time mechanism for threat detection, conformity assessment, or asset management can deal with local security problems and systems uncertainties.

The threat profiling mechanism exploits the system's characteristics like service availability, average request latency, failure rate, and raw data collected from the system logs, events, and processes running on the EG-IG to generate more concrete information. The profiles use a tree data structure for describing clients' activities and the system's characteristics. These threat profiles can be used by strategic applications to determine inherent weaknesses, identify possible threats, and predict zero-day cyber-attack scenarios that can adversely impact the systems.

\subsection{Edge-Computing Systems}
Edge-computing systems for business intelligence, intelligent transportation, industrial plants, and smart cities conception are gaining traction. Multi-Access Edge Computing (MEC) can aid in network congestion minimization, resource optimization, and user experience enhancement by reducing the traffic propagating through the different levels of the automation pyramid, lowering the latency for applications and services, and scaling up remote maintenance services.  

However, the edge devices' trustworthiness is essential for the distributed applications to use them reliably. There are open challenges, such as programmability, better data abstractions and services for building adaptive applications, debugging and testing of edge-site applications, interoperability of heterogeneous communication technologies, transport layer security from Denial of Service (DoS)/Distributed DoS (DDoS) attacks, communication attacks like message forging, message tampering, or reply attack require further investigation.

\section{Conclusions}\label{sec:Conclusions}
We designed and developed an EC-IG for Industry 4.0 by leveraging Trustzone technology for IT/OT convergence that can facilitate secure vertical and horizontal integration. Further, as a proof-of-concept, we implemented a remote production-line maintenance use case for diagnostic management, firmware management, control-level management, and threat profiles management. Security by isolation using Trustzone technology provided a trusted edge platform supporting secure computation, secure storage, and remote attestation, thus, guaranteeing the CIA security properties.

Our future work will emphasize the security assessment of the EC-IG prototype. We will evaluate the prototype against both active (e.g., unauthorized access, spoofing, privilege escalation, denial of services) and passive (e.g., information leakage, eavesdropping, network, and traffic analysis) cyber attacks that can disrupt EC-IG operations. We will also expand the threat profiling concept by establishing theories and principles to perform predictive cyber risk analysis for resource-limited edge devices.
\bibliographystyle{elsarticle-num}
\bibliography{main}

\end{document}